\documentclass[bibyear]{aa} 

\usepackage{graphicx}
\usepackage{txfonts}
\begin{document}

   \title{IDENTIFYING AGN's BALMER-ABSORPTIONS AND STRATIFIED NLR KINEMATICS IN SDSS\,J112611.63+425246.4}


   \author{J. Wang
          \inst{1}
          \and
          D. W. Xu\inst{1}
          }

  \institute{National Astronomical Observatories, Chinese Academy of Sciences\\
              \email{wj@nao.cas.cn}
             }


 
  \abstract
   {Balmer absorption is a rare phenomenon in active galactic nuclei (AGNs). So far, only seven Balmer-absorption AGNs have been reported in literature.}
   {We here report the identification of SDSS\,J112611.63+425246 as a new Balmer-absorption AGN through our spectral analysis, and study the
   kinematics of its narrow emission-line region (NLR).}
   {We model the continuum by a linear combination of a starlight component, a powerlaw from the central AGN and the emission from the FeII complex. 
     After the subtraction of the modeled continuum, each emission/absorption line is profiled by a sum of multi Gaussian functions. All the line 
    shifts are determined with respect to the modeled starlight component.}
   {By using the host starlight as a reference for the local system, both H$\alpha$ and H$\beta$ show AGN's absorptions with 
    a blueshift of $\sim300\mathrm{km\ s^{-1}}$. We identify a strong anti-correlation between the inferred velocity shifts and 
    ionization potential for various narrow emission lines, which suggests a stratified NLR kinematics.  
    A de-accelerated outflow is implied for the inner NLR gas, while an accelerated inflow for the outer NLR gas.
    This complicated NLR kinematics additionally implies that AGN's narrow emission lines, even for the low-ionized lines, 
    might not be a reliable surrogate for the velocity of the local system.

}
   {}

   \authorrunning{J. Wang \& D. W. Xu}
   \titlerunning{BALMER-ABSORPTIONS AND STRATIFIED NLR KINEMATICS IN SDSS\,J112611.63+425246.4}
   \maketitle

%

\section{Introduction}

The feedback from a  central active galactic nucleus (AGN) is now believed to be
a potential key ingredient in understanding the coevolution of the AGN and its
host galaxy. A widely accepted scenario is that
the growth of supermassive black hole (SMBH) regulates host star formation by sweeping out
circumnuclear gas (e.g., Silk \& Rees 1998; Somerville et al. 2008; Hirschmann et al. 2013;
Di Matteo et al. 2007; Fabian 1999;  Granato et al. 2004; Croton et al. 2006).

The evidence of outflow from an AGN
in various scales could be identified in multi-wavelength bands from radio to X-ray (see Veilleux et al. 2005 and
Fabian 2012 for reviews).
AGN's outflow could be diagnosed by the blueshifted absorption lines.
Although the narrow absorption lines with width $\leq 500\mathrm{km\ s^{-1}}$ are
frequently identified in type I AGNs in UV and X-ray  ($\sim50\%$, e.g.,  Hamann \& Sabra 2004), Balmer-absorption
AGNs are still rare cases. So far, only seven Balmer-absorption AGNs are reported in literature, they are: NGC\,4151 (Hutching et al. 2002),
SDSS\,J0839+3805 (Aoki et al. 2006), SDSS\,J1259+1213 (Hall 2007), SDSS\,J1029+4500 (Wang et al. 2008),
SDSS\,1723+5553 (Aoki 2010), LBQS\,1206+1052 (Ji et al. 2012), and SDSS\,J2220+0109 (Ji et al. 2013). Because of
their rarity, the identification of more Balmer-absorption AGNs is therefore essential for subsequent study of the nature of
the AGN's Balmer absorption-line region (BAR). At first, recent studies point out that rigorous condition with high hydrogen 
column density of $\sim10^{21-22}\mathrm{cm^{-2}}$ is required to excite neutral hydrogen atoms to  
$n=2$ shell by Ly$\alpha$ resonant pumping (e.g., Ji et al. 2012).  
Secondly, the Balmer absorption lines can be used as a diagnostic for the kinematics of the natural gas around central AGNs.

In this paper, we report a detailed spectroscopic analysis for SDSS\,J112611.63+425246.4 
(hereafter SDSS\,J1126+4252 for short)\footnote{This object has been analyzed by Hu et al. (2008) 
in their large type-I AGN sample. The Balmer absorptions and starlight component were, however, not taken into account 
in their spectral modelings.}, which allows us to
identify the object as a new Balmer-absorption AGN and to identify a stratified kinematics in its narrow 
emission-line region (NLR) with respect to the systematic velocity determined from the host galaxy.


\section{Spectral Analysis}

The optical spectrum of SDSS\,J1126+4252 is extracted when we carried out a systematic
X-ray and optical spectral analysis on the XMM-Newton 2XMMi/SDSS-DR7 catalog that is
originally crossmatched by Pineau et al (2011). The catalog contains
a total of more than 30,000 X-ray point-like sources (with an X-ray
position accuracy $\leq 5\arcsec$) that have a SDSS-DR7 optical counterpart
with an identification probability larger than 90\%. The spectrum of the object 
was taken by the SDSS dedicated 2.5m wide field telescope in February 27th, 2004.

The 1-Dimensional spectrum of the object is analyzed by the IRAF\footnote{IRAF is distributed by National Optical Astronomy
Observatory, which is operated by the Association of Universities
for Research in Astronomy, Inc., under cooperative agreement with the National Science Foundation.} package,
including Galactic extinction correction, transformation to the rest
frame, starlight component removal and emission/absorption line profiling.
We at first correct the Galactic extinction by the color excess $E(B-V)$ taken from the Schlegel,
Finkbeiner, and Davies Galactic reddening map (Schlegel et al. 1998), by assuming an $R_V$ = 3.1
extinction law of the Milky Way (Cardelli et al. 1989). The spectrum is then de-redshifted to its
rest frame, along with the flux correction due to the relativity effect basing upon the measured redshift
provided by the SDSS pipelines. The object has a nominal redshift of $z=0.15592\pm0.00121$, which 
corresponds to a velocity uncertainty of $363\mathrm{km\ s^{-1}}$.

The total light spectrum at the rest frame is displayed in Figure 1. It
shows that there is non-negligible contamination from its host galaxy. To isolate the AGN's emission-line spectrum, 
we model the continuum by a linear
combination of a powerlaw continuum, an FeII complex template and the seven eigenspectra of starlight. 
The adopted FeII template is taken from Boroson \& Green (1992). The eigenspectra were
built from the standard single stellar
population spectral library developed by Bruzual \& Charlot
(2003) through the principal component analysis (PCA) method (e.g., Wang \& Wei 2008;
Francis et al. 1992). An additional Galactic extinction curve with $R_V=3.1$ is included in the modeling to account for the
intrinsic extinction of the host galaxy. $\chi^2$ minimizations are iteratively performed over the rest-frame wavelength
range from 3700 to 7000\AA, except for the regions with strong emission lines. The line width of the FeII 
template and the velocity dispersion of the starlight are pre-determined through cross-correlation method before each iteration. \rm
The removal of the continuum is illustrated in Figure 1 as well.



The AGN's emission/absorption lines are subsequently modeled on the isolated line spectrum
by using the SPECFIT task (Kriss 1994) in the IRAF package. The line modelings 
are schemed in the left and right panels in Figure 2 for the H$\alpha$ and H$\beta$ regions, respectively.
The two narrow Balmer absorptions are marked by the vertical short lines in the figure. Each emission line is modeled by a
free Gaussian function, except for the broad H$\alpha$ emission and the [OIII] doublet. 
Figure 2 clearly shows that a linear combination of two broad H$\alpha$ components are required to properly reproduce the observed
line profile. We measure the FWMH of the total broad H$\alpha$ emission by the IRAF/SPLOT task from the
residual profile that is obtained by subtracting the modeled narrow emission (i.e., H$\alpha$, [NII]$\lambda\lambda$6548,6583) 
and absorption lines from the the observed profile. In addition to the narrow peak, a broad and blueshift component
is necessary to model the slight blue wing of the [OIII]$\lambda5007$ profile. The [OIII]$\lambda4959$ ([NII]$\lambda$6548) line 
profile is set to be the same as [OIII]$\lambda5007$ ([NII]$\lambda$6583). 
The intensity ratio of the [OIII] ([NII]) doublet is fixed to the theoretical value of 3.
In total, the freedom in the $\chi^2$ minimization is 21 and 15 for the H$\alpha$ and H$\beta$ regions, respectively.

\section{Results and Discussions}

The measured line properties are tabulated in Table 1.
The reported flux of the H$\alpha$ broad emission (and the [OIII]$\lambda$5007 line emission) is the sum of the 
two fitted components. The quoted line width and velocity shift is based on the fitted narrow peak for the [OIII] line.
The flux of the FeII blends (FeII$\lambda4570$) is measured in the rest-frame wavelength range from 4434 to
4684\AA, which results in a parameter of RFe of $0.64\pm0.20$. RFe is defined as
the flux ratio between the FeII$\lambda4570$ and H$\beta$ broad component. 
All the reported line widths
are not corrected for the intrinsic instrument resolution of $\sigma_{\mathrm{inst}}\approx65\mathrm{km\ s^{-1}}$.
Thanks to the evident contamination of the starlight in the integrated spectrum, we emphasize that  
the reported line shifts are all calculated with respect to the modeled starlight component\footnote{Hu et al. (2008) uses the [OIII]$\lambda5007$ line as 
a reference, and find that the [OII] emission line might be a more reliable reference than either [OIII] or H$\beta$. Our results are 
consistent with their measurements if the [OIII] line is used as a reference. }: $\Delta\upsilon=\Delta\upsilon_{\mathrm{line}}-\Delta\upsilon_{\mathrm{host}}$,
where $\Delta\upsilon_{\mathrm{line}}$ and $\Delta\upsilon_{\mathrm{host}}$ are the modeled velocity shifts with respect to the 
nominal redshift for a given emission/absorption line and for the host galaxy, respectively.
A negative value of $\Delta\upsilon$ corresponds to a blueshift, and a positive one to a redshift.

 All the uncertainties reported in the table (except for the FWHM of H$\alpha_{\mathrm{b}}$) only include the 
errors at 1$\sigma$ significance level resulted from the $\chi^2$ 
minimizations. The error of the FWHM of H$\alpha_{\mathrm{b}}$ is obtained by a statistic on the multiple measurements 
by the IRAF/SPLOT task. A proper error propagation is taken into account in the reported uncertainties of $\Delta\upsilon$.

\subsection{Balmer absorption lines}

We argue that the observed blueshifted Balmer absorption lines are most likely resulted from an outflow from
central engine, rather than from the host galaxy. In fact, the starlight component has been properly removed
from the observed integral spectrum as described above. Moreover, the lack of a strong Balmer break enables us to
exclude the case in which the observed Balmer absorptions are from a post-starburst galaxy with
strong Balmer absorptions (e.g., Brotherton et al, 1999; Wang \& Wei 2006). In interpretation of the Balmer
absorptions in AGNs, the intrinsic EW depends on whether the absorbing gas covers the BLR or not (e.g, de Kool et al. 2001).
The later scenario is favored in the object by comparing the measured Balmer absorption EW ratio to its theoretical
value. In an absorption line without saturation, its EW could be related to its column density $N$ as (Jenkins 1986):

\begin{equation}
 EW_\lambda=\frac{\pi e^2f\lambda^2N}{m_ec^2}
\end{equation}
where $f$ is the oscillator strength. A theoretical $f\lambda$ value of 7.26 is therefore expected for the H$\alpha$ to H$\beta$
ratio, which is very close to the observed EW ratio of $EW(\mathrm{H}\alpha)/EW(\mathrm{H}\beta)=7.38\pm4.32$ when both absorption lines are normalized with
respective to the modeled AGN's continuum. On the contrary, the observed ratio is closed to 1 if both absorption lines are normalized to the corresponding broad emission
line. This comparison therefore allows us to believe that the absorbing gas being responsible for the Balmer transitions is not saturated and fully covers the continuum source.

We further estimate the neutral hydrogen column density from Eq. (1). The inferred column densities of hydrogen at $n=2$ shell from the H$\alpha$ and H$\beta$
absorption lines are
$N_{\mathrm{HI,2}} = (1.2\pm0.5)\times10^{14}\ \mathrm{cm^{-2}}$ and $(1.6\pm2.2)\times10^{14}\ \mathrm{cm^{-2}}$, respectively.
The neutral hydrogen column density could be derived from $N_{\mathrm{HI}}\approx N_{\mathrm{HI,1}}+N_{\mathrm{HI,2}}$, where $N_{\mathrm{HI,1}}$ is the column
density of hydrogen at $n=1$ shell and is estimated by following Hall (2007):
\begin{equation}
 \frac{N_{\mathrm{HI,1}}}{N_{\mathrm{HI,2}}}=\frac{1}{4\tau_{\mathrm{Ly\alpha}}}e^{\frac{10.2\mathrm{eV}}{kT}}
\end{equation}
where $\tau_{\mathrm{Ly\alpha}}$ is the optical depth at the center of the Ly$\alpha$ absorption. The depth $\tau_{\mathrm{Ly\alpha}}$ could be
inferred from the relationship $\tau_{\mathrm{Ly}\alpha}=0.12\tau_{\mathrm{H\alpha}}(N_{\mathrm{HI,1}}/N_{\mathrm{HI,2}})$ (see Eq. (1) in Aoki 2010).
Substituting this relationship into Eq (2) results in a relation
\begin{equation}
 \frac{N_{\mathrm{HI,1}}}{N_{\mathrm{HI,2}}}=\frac{1.44}{\sqrt{\tau_{\mathrm{H\alpha}}}}e^{\frac{5.1\mathrm{eV}}{kT}}
\end{equation}
Taking $T=7500$K (Osterbrock \& Ferland 2006) and $\tau_{\mathrm{H\alpha}}=1.69\times10^5EW(\mathrm{H\alpha})/\lambda/b=8.66$ (where $b$ is the Doppler parameter of the absorption line after the correction
of the intrinsic instrumental resolution), the inferred neutral hydrogen column density is $\sim1.5\times10^{17}\ \mathrm{cm^{-2}}$.

\subsection{UV and X-ray observations}

SDSS\,J1126+4252 is particular weak and hard in UV and X-ray bands. The object is a common source in both second \it XMM-Newton \rm serendipitous source
catalog (XMMSSC) and \it XMM-Newton \rm optical Monitor serendipitous UV source survey catalog (XMMOMSUSS).
Vagnetti et al. (2010) shows that
the inferred specific
luminosities at $2500\AA$ and 2keV are as low as $4.8\times10^{28}\mathrm{erg\ s^{-1}\ Hz^{-1}}$ and $6.6\times10^{23}\mathrm{erg\ s^{-1}\ Hz^{-1}}$,
respectively. Its very hard X-ray spectrum could be additionally  learned from the very large hardness ratios\footnote{The hardness ratios
are defined as $HR3=(CR4-CR3)/(CR4+CR3)$ and $HR4=(CR5-CR4)/(CR5+CR4$, where
$CR3$, $CR4$ and $CR5$ are the count rates in the energy bands 1-2, 2-4.5,4.5-12keV, respectively.}: $HR3=0.52$ and $HR4=0.63$.

Ji et al. (2012, 2013) recently point out that a rigorous condition is required for the formation
of Balmer absorptions. The absorptions are likely caused by Ly$\alpha$ resonant pumping in a partially ionized region with 
a high column density of $N_{\mathrm{H}}\sim10^{21-22}\ \mathrm{cm^{-2}}$.
A heavy obscuration due to the required high column density is a possible explanation of the observed  extremely weak and hard emission in both UV and X-ray.

\subsection{Eddington ratio and SMBH mass}

We estimate the SMBH mass $M_{\mathrm{BH}}$ in terms of its H$\alpha$ broad component
according to the calibration in Greene \& Ho (2007):
\begin{equation}
M_{\mathrm{BH}}=3.0^{+0.6}_{-0.5}\times10^6\bigg(\frac{L_{\mathrm{H\alpha}}}{10^{42}\ \mathrm{erg\ s^{-1}}}\bigg)^{0.45\pm0.03}
\bigg(\frac{\mathrm{FWHM_{H\alpha}}}{1000\ \mathrm{km\ s^{-1}}}\bigg)^{2.06\pm0.06}\ M_\odot
\end{equation}
where $L_{\mathrm{H\alpha}}$ is the intrinsic luminosity of the H$\alpha$ broad component
corrected for the local extinction.
The extinction is inferred from the narrow-line ratio H$\alpha$/H$\beta$, assuming
the Balmer decrement for standard case B recombination and a Galactic extinction
curve with $R_V = 3.1$. With the estimated $M_{\mathrm{BH}}$, the Eddington ratio $L/L_{\mathrm{Edd}}$
(where $L_{\mathrm{Edd}}=1.26\times10^{38}M_{\mathrm{BH}}/M_\odot$ is
the Eddington luminosity) is obtained from a combination of the bolometric correction of $L_{\mathrm{bol}}=9\lambda L_{\lambda}(5100\AA)$ (Kaspi et al. 2000)
and the $L_{5100\AA}-L_{\mathrm{H\alpha}}$ relation reported in Greene \& Ho (2005)
\begin{equation}
\lambda L_{\lambda}(5100\AA)=2.4\times10^{43}\bigg(\frac{L_{\mathrm{H\alpha}}}{10^{42}\ \mathrm{erg\ s^{-1}}}\bigg)^{0.86}\ \mathrm{erg\ s^{-1}} 
\end{equation}
This luminosity relation has a rms scatter of 0.2dex around the best-fit line. 
The calculated $M_{\mathrm{BH}}$ and $L/L_{\mathrm{Edd}}$ are $\approx1.8\times10^8\ M_\odot$ and $\approx0.06$, respectively.
By combining the intrinsic scatters of the used relationships and the uncertainties derived from our line modelings, a proper error 
propagation returns 1$\sigma$ uncertainties of 0.40dex and 0.45dex for the calculated $M_{\mathrm{BH}}$ and $L/L_{\mathrm{Edd}}$.

We argue that the inferred $M_{\mathrm{BH}}$ from the broad H$\alpha$ emission agrees with the properties of the host galaxy.
With the velocity dispersion of the host galaxy of $\sigma_\star\sim270\mathrm{km\ s^{-1}}$ obtained from our continuum modeling,
the $M_{\mathrm{BH}}-\sigma_\star$ relation of $\log(M_{\mathrm{BH}}/M_\odot)=8.13+4.02\log(\sigma_\star/200\mathrm{km\ s^{-1}})$ 
(Tremaine et al. 2004) yields a blackhole mass of $\log(M_{\mathrm{BH}}/M_\odot)\sim8.7$, which is highly consistent with 
the value estimated from the broad H$\alpha$ emission.

\subsection{Stratified NLR kinematics}

The spectral analysis allows us to study the line shifts in SDSS\,J1126+4252 by using its host starlight component as 
a reference of the systematic velocity. One can see 
from Table 2 that all the low-ionized narrow emission lines show a redshift with respect to its 
host galaxy, while a blueshift could be identified in the high-ionized emission line [NeIII]$\lambda3868$. 
It is interesting that the [OIII]$\lambda5007$ emission line has a marginal
blueshift of $\Delta\upsilon=-10\pm30\mathrm{km\ s^{-1}}$.

A strong anti-correlation between the velocity shifts and ionization potential (IP) is shown in Figure 3.
The velocity shift of the emitting gas of neutral hydrogen atom is taken from the measurement of narrow H$\alpha$ emission, 
both because the narrow H$\alpha$ and H$\beta$ emission show comparable velocity shifts and because of 
the higher signal-to-noise ratio of the narrow H$\alpha$ line.
An average value of velocity shift is adopted in the figure for the [SII] doublet.  
The best fit yields a relation of $\Delta\upsilon=(186.4\pm35.4)-(4.00\pm0.80)\mathrm{IP}$.  In fact, 
Komossa et al. (2008) propose a similar correlation between the line shift and IP
in the narrow-line Seyfert 1 galaxies with large [OIII] line blueshift over $150\mathrm{km\ s^{-1}}$.

Because the AGN's NLR gas is believed to be generally stratified in density and ionization potential 
(e.g., Filippenko \& Halpern 1984; Filippenko 1985; De Robertis \& Osterbrock 1986),
our fitted relationship implies a complicate NLR kinematics in SDSS\,J1126+4252. 
A de-accelerated outflow is expected for the inner NLR gas, while an accelerated inflow for 
the outer NLR gas. The turnover of the radial velocity occurs at the [OIII]$\lambda5007$ emission-line gas
whose radial velocity shift with respect to the local system determined from the host starlight is very close to zero. \rm
Although the outflows from central AGNs in various scales have been frequently identified in AGNs 
(e.g., Komossa et al. 2008 and Fabian 2012 for a recent review), the inflows have already been revealed in 
a few of nearby AGNs through integrated field spectroscopic observations in both 
optical band and near-infrared (e.g., Fathi et al. 2005; Storchi-Bergmann et al. 2007; Riffel et al. 2008, 2013;
Riffel \& Storchi-Bergmann 2011). The observations reveal distinct kinematics for different emitting gas.
The inflowing gas to the central active nucleus could be traced by the $\mathrm{H_2}$ emission, and the outflowing 
gas by the [FeII] emission, which is similar to the kinematics revealed in SDSS\,J1126+4252.

The implications described above is based on the scenario in which the detected NLR lines are seen in front of 
the central AGN. We here can not exclude an alternatively scenario in which these NLR lines are seen behind 
the central source. In this scenario, the obtained relationship implies a de-accelerated inflow for the inner NLR gas, 
and an accelerated outflow for the outer NLR gas.

Our spectral analysis indicates that in SDSS\,J1126+4252 all the narrow emission lines, except for the [OIII], show non-negligible velocity 
shifts with respect to the local system determined from the host starlight.
This fact puts forward a worry that AGN's narrow emission lines, even for the low-ionized lines, might be 
a non-reliable surrogate for the velocity of the local system. A large sample 
is needed to perform a further examination on the relationship between the velocity shifts of various
narrow emission-lines and local system determined from host starlight in
the feature.   

   \begin{figure}
   \centering
   \includegraphics[width=8cm]{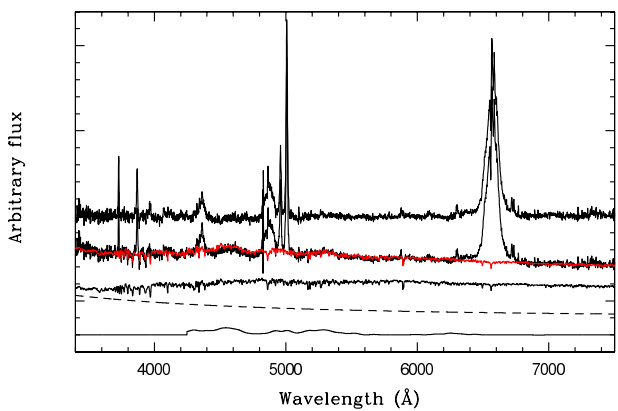}
   \caption{Illustration of the modeling and removal of the continuum. The continuum subtracted emission-line spectrum is 
shown by the top curve. Below the emission-line spectrum, the modeled continuum is over plotted by the red curve on the observed spectrum.
The modeled continuum is obtained by a linear combination of a starlight component, a powerlaw continuum from the AGN and the emission from
the FeII complex, which are displayed in ordinal below the observed spectrum.
All the spectra are shifted vertically by an arbitrary amount for visibility.}
              \label{Fig1}%
    \end{figure}
   \begin{figure}
   \centering
   \includegraphics[width=8cm]{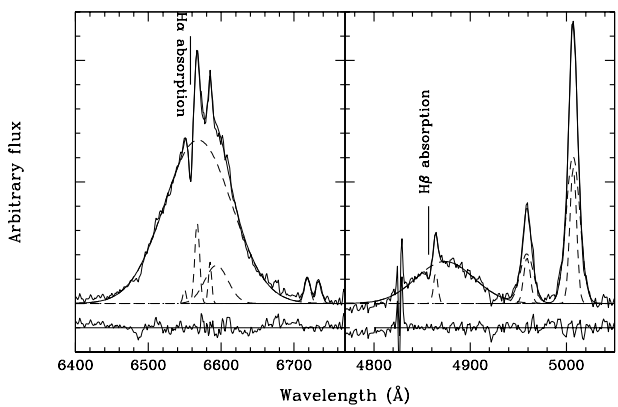}
   \caption{Line profile modelings by a sum of a set of Gaussian components for the H$\alpha$ (left panel) and
H$\beta$ (right panel) regions.  The blueshifted Balmer absorptions are marked by the short vertical lines.
In each panel, the observed and modeled line profiles are plotted
by the light and heavy solid lines, respectively. Each Gaussian component in emission is
shown by a dashed line. The sub-panel underneath each line spectrum
presents the residuals between the observed and modeled profiles.}
              \label{Fig2}%
    \end{figure}
  
   \begin{figure}
   \centering
   \includegraphics[width=8cm]{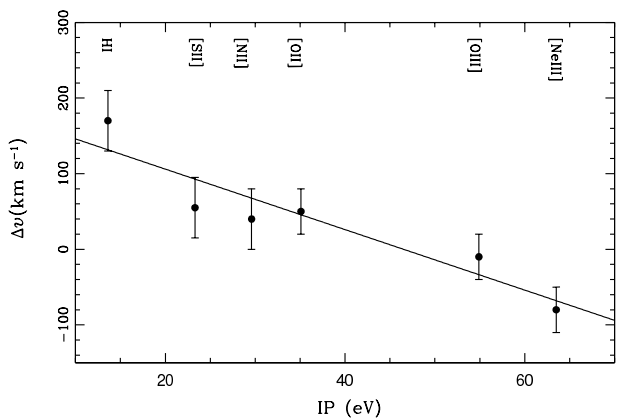}
   \caption{Correlation between velocity shifts and ionization potential (IP). A positive velocity denotes a redshift, and 
   a negative one a blueshift. The best fit line is over plotted by the solid line.}
              \label{Fig2}%
    \end{figure}

\begin{table}
\caption{Spectral properties of SDSS\,J1126+4252}             
\label{table:1}      
\centering                          
\begin{tabular}{l c c c}        
\hline\hline 
\multicolumn{4}{c}{Emission}\\ 
\hline
Line  & Flux & FWHM & $\Delta\upsilon$ \\  
                    & $10^{-15}\mathrm{erg\ s^{-1}\ cm^{-2}}$ & $\mathrm{km\ s^{-1}}$ & $\mathrm{km\ s^{-1}}$\\
(1) & (2) & (3) & (4)\\
\hline
$\mathrm{[OII]}\lambda$3727\dotfill & $2.3\pm0.2$ & $520\pm40$  &  $50\pm30$ \\
$\mathrm{[NeIII]}\lambda$3868 &  $2.6\pm0.2$ &  $710\pm50$ & $-80\pm30$\\ 
$\mathrm{FeII\lambda4570}$\dotfill & $8.3\pm0.3$ & $\sim3\times10^3$ & $100\pm150$\\
H$\beta_n$\dotfill & $0.7\pm0.5$     &   $340\pm250$ & $150\pm80$ \\
H$\beta_b$\dotfill & $13.0\pm1.1$ &   $4350\pm230$ & $750\pm30$\\
$\mathrm{[OIII]}\lambda$5007\dotfill & $15.8\pm2.9$ & $500\pm90$ & $-10\pm30$\\
H$\alpha_n$\dotfill & $2.7\pm1.1$ & $350\pm110$ & $170\pm40$\\
H$\alpha_b$\dotfill & $84.3\pm2.8$ & $4640.0\pm70$ & \dotfill\\
$\mathrm{[NII]}\lambda$6583\dotfill & $1.0\pm0.6$ & $240\pm120$ & $40\pm40$\\
$\mathrm{[SII]\lambda6716}$\dotfill & $1.1\pm0.1$ & $440\pm40$ & $50\pm30$\\
$\mathrm{[SII]\lambda6731}$\dotfill & $1.0\pm0.1$ & $440\pm40$ & $60\pm30$\\
\hline
\multicolumn{4}{c}{Absorption}\\
\hline
H$\beta$\dotfill & $28.8\pm11.1$ &   $290\pm540$ & $-300\pm70$\\
H$\alpha$\dotfill & $3.9\pm5.6$     &   $210\pm70$ & $-250\pm40$\\
\hline                                   
\end{tabular}
\end{table}

%

%

%

\section{Conclusions}
Detailed spectral analysis is performed on SDSS\,J11261.63+425246.4, which allows us to identify the object as a new 
Balmer-absorption AGN. Using the modeled host starlight as the reference of the local system, a stratified kinematics is
identified in the NLR of the object, i.e., a strong anti-correlation between 
the inferred velocity shifts and ionization potentials. The revealed relationship
implies a de-accelerated outflow stream for its inner NLR gas, and an accelerated inflow stream for its outer NLR gas.

\begin{acknowledgements}
The authors thank Profs. Todd A.
Boroson and Richard F. Green for providing the optical FeII
template. This research has made use of the SDSS archive data that are created and
distributed by the Alfred P. Sloan Foundation. JW 
is supported by the National Natural Science Foundation of
China (grant No. 11473036). DWX is supported by the National Natural Science Foundation of
China under grant No. 11273027.
\end{acknowledgements}



\begin{thebibliography}{}
\bibitem[Aoki 2010]{aok10} Aoki, K.  2010, \pasj, 62, 1333
\bibitem[Aoki et al. 2006]{aok06} Aoki, K., Iwata, I., Ohta, K., Ando, M., Akiyama, M., \& Tamura, N. 2006, \apj, 651, 84
\bibitem[Boroson \& Green 1992]{bog92} Boroson, T. A., \& Green, R. F. 1992, \apjs, 80, 109
\bibitem[Brotherton et al. 1999]{bro99} Brotherton, M. S., et al. 1999, \apjl, 520, 87
\bibitem[Bruzual \& Charlot 2003]{brc03} Bruzual, G., \& Charlot, S. 2003, \mnras, 344, 1000
\bibitem[Cardelli et al. 1989]{car89} Cardelli, J. A., Clayton, G. C., \& Mathis, J. S. 1989, \apj, 345, 245
\bibitem[Croton et al. 2004]{cro04} Croton, D. J., Springel, V., White, S. D. M., et al. 2006, \mnras, 365, 11
\bibitem[de Kool et al. 2001]{dek01} de Kool, M., Arav, N., Becker, R. H., et al. 2001, \apj, 548, 609
\bibitem[De Robertis \& Osterbrock 1986]{deo86} De Robertis, M. M., \&Osterbrock, D. E. 1986, ApJ, 301, 727	
\bibitem[Di Matteo et al. 2007]{dim07} Di Matteo, P., Combes, F., Melchior, A.-L., \& Semelin, B. 2007, \aap, 468, 61
\bibitem[Fabian 1999]{fab99} Fabian, A. C. 1999, \mnras, 308, L39
\bibitem[Fabian 2012]{fab12} Fabian, A. C. 2012, \araa, 50, 455
\bibitem[Fathi et al. 2006]{fat06} Fathi, K., et al. 2006, ApJL, 641, 25
\bibitem[Filippenko 1985]{fil85} Filippenko, A. V. 1985, ApJ, 289, 475
\bibitem[Filippenko \& Halpern 1985]{fih85} Filippenko, A. V. \& Halpern, J. P. 1984, ApJ, 285, 458
\bibitem[Francis et al. 1992]{fra92} Francis, P. J., Hewett, P. C., Foltz, C. B., \& Chaffee, F. H. 1992, \apj, 398, 476
\bibitem[Granato et al. 2004]{gra04} Granato, G. L., De Zotti, G., Silva, L., Bressan, A., \& Danese, L. 2004, \apj, 600, 580
\bibitem[Greene \& Ho 2005]{grh05} Greene, J. E., \& Ho, L. C. 2005, \apj, 630 ,122
\bibitem[Greene \& Ho 2007]{grh07} Greene, J. E., \& Ho, L. C. 2007, \apj, 670 ,92
\bibitem[Hall 2007]{hal07} Hall, P. B. 2007, \aj, 133, 1271
\bibitem[Hamann \& Sabra 2004]{has04} Hamann, F., \& Sabra, B. 2004, ASPC, 311, 203
\bibitem[Hu et al. 2008]{nu08} 	Hu, C., Wang, J. M., Ho, L. C., Chen, Y. M., Zhang, H, T., Bian, W. H., \& Xue, S. J. 2008, ApJ, 687, 78
\bibitem[Hutchings et al. 2002]{hut02}  Hutchings, J. B., Crenshaw, D. M., Kraemer, S. B., Gabel, J. R., Kaiser, M. E., Weistrop, D., \& Gull, T. R. 2002, \aj, 124, 2543
\bibitem[Hirschmann et al. 2013]{hir13} Hirschmann, M., et al. 2013, \mnras, 436, 2929
\bibitem[Jenkins 1986]{jen86} Jenkins, E. B. 1986, \apj, 304, 739
\bibitem[Ji et al. 2012]{ji12} Ji, T., Wang, T -G., Zhou, H -Y., \& Wang, H -Y. 2012, RAA, 12, 369
\bibitem[Ji et al. 2013]{ji13} Ji, T., Zhou, H -Y., Wang, T -G., \& Wang, H -Y. 2013, ChA\&A, 37, 17
\bibitem[Kaspi et al. 2000]{kas00} Kaspi, S., et al. 2000, ApJ, 533, 631
\bibitem[Komossa et al. 2008]{kom08} Komossa, S., Xu, D., Zhou, H., \& Storchi-Bergmann, T. 2008, ApJ, 680, 926
\bibitem[Kriss 1994]{kri94} Kriss, G. 1994, Adass, 3, 43
\bibitem[Osterbrock \& Ferland 2006]{osf06} Osterbrock, D. E., \& Ferland, G. J. 2006, Astrophysics of Gaseous Nebulae and Active Galactic Nuclei, 2nd edition.
\bibitem[Pineau et al. 2011]{pin11} Pineau, F.-X., Motch, C., Carrera, F., et al. 2011, \aap, 527, 126
\bibitem[Riffel \& Storchi-Bergmann 2013]{rif13} Riffel, R. A., \& Storchi-Bergmann, T. 2011, MNRAS, 417, 2752
\bibitem[Riffel et al. 2013]{rif08} Riffel, R. A., Storchi-Bergmann, T., Winge, C. 2013, MNRAS, 430, 2249
\bibitem[Riffel et al. 2008]{rif08} Riffel, R. A., Storchi-Bergmann, T., Winge, C., McGregor, P. J., Beck, T., \& Schmitt, H. 2008, MNRAS, 385, 1129
\bibitem[Schlegel et al. 1998]{sch98} Schlegel, D., Finkbeiner, D. P., \& Davis, M. 1998, \apj, 500, 525
\bibitem[Silk \& Rees 1998]{sir98}Silk, J., \& Rees, M. J. 1998, \aap, 331, L1
\bibitem[Somerville et al. 2008]{som08} Somerville, R. S., Hopkins, P. F., Cox, T. J., Robertson, B. E., \& Hernquist, L. 2008, \mnras, 391, 481
\bibitem[Storchi-Bergmann et al. 2007]{sto07} Storchi-Bergmann, T., et al. 2007, ApJ, 670, 959
\bibitem[Tremaine et al. 2004]{tre04} Tremaine, C. A., Heckman, T. M., Kauffmann, G., et al. 2004, ApJ, 613, 898
\bibitem[Vagnett\i et al. 2010]{vag10} Vagnetti, F., Turriziani, S., Trevese, D., \& Antonucci, M. 2010, \aap, 519, 7
\bibitem[Veilleux et al. 2005]{vei05} Veilleux, S., Cecil, G., \& Bland-Hawthorn, J. 2005, ARA\&A, 43, 769
\bibitem[Wang \& Wei 2006]{waw06} Wang, J., \& Wei, J. Y. 2006, \apj, 648, 158
\bibitem[Wang \& Wei 2008]{waw08} Wang, J., \& Wei, J. Y. 2008, \apj, 679, 86
\bibitem[Wang et al. 2008]{wan08} Wang, T., Dai, H., \& Zhou, H. 2008, \apj, 674, 668
\end{thebibliography}
\end{document}